\documentclass[twocolumn,reprint,superscriptaddress,aps,floatfix,showkeys]{revtex4-2}

\usepackage{graphicx}
\usepackage{amsmath,amssymb,amsthm,mathrsfs,amsfonts,dsfont,amstext}
\usepackage{textcomp,pbox}
\usepackage[colorlinks=true]{hyperref}
\usepackage{graphicx}
\usepackage{dcolumn,booktabs,url}
\usepackage{bm}
\usepackage{color}
\usepackage{latexsym}

\newcommand{\figref}[1]{\figurename{~\ref{#1}}}

\hypersetup{
     colorlinks   = true,
     citecolor    = blue,
     linkcolor    = blue,
     urlcolor     = blue     
}

\begin{document}
\title{Eliminating temporal correlation in quantum-dot entangled photon source by quantum interference}

\newcommand{\USTCmicro}{Hefei National Research Center for Physical Sciences at the Microscale and School of Physical Sciences, University of Science and Technology of China, Hefei 230026, China}
\newcommand{\USTCCAS}{Shanghai Research Center for Quantum Science and CAS Center for Excellence in Quantum Information and Quantum Physics, University of Science and Technology of China, Shanghai 201315, China}
\newcommand{\USTCNational}{Hefei National Laboratory, University of Science and Technology of China, Hefei 230088, China}
\author{Run-Ze Liu}
\thanks{These authors contributed equally to this work.}
\affiliation{\USTCmicro}
\affiliation{\USTCCAS}
\author{Yu-Kun Qiao}
\thanks{These authors contributed equally to this work.}
\affiliation{\USTCmicro}
\affiliation{\USTCCAS}
\author{Han-Sen Zhong}
\thanks{These authors contributed equally to this work.}
\affiliation{\USTCmicro}
\affiliation{\USTCCAS}
\author{Zhen-Xuan Ge}
\affiliation{\USTCmicro}
\affiliation{\USTCCAS}
\author{Hui Wang}
\affiliation{\USTCmicro}
\affiliation{\USTCCAS}
\author{Tung-Hsun Chung}
\affiliation{\USTCNational}
\author{Chao-Yang Lu}
\affiliation{\USTCmicro}
\affiliation{\USTCCAS}
\affiliation{\USTCNational}
\author{Yong-Heng Huo}
\email{yongheng@ustc.edu.cn}
\affiliation{\USTCmicro}
\affiliation{\USTCCAS}
\affiliation{\USTCNational}
\author{Jian-Wei Pan}
\affiliation{\USTCmicro}
\affiliation{\USTCCAS}
\affiliation{\USTCNational}

\date{\today}
\begin{abstract}
Semiconductor quantum dots, as promising solid-state platform, have exhibited deterministic photon pair generation with high polarization entanglement f\textcompwordmark idelity for quantum information applications. However, due to temporal correlation from inherently cascaded emission, photon indistinguishability is limited, which restricts their potential scalability to multi-photon experiments. Here, by utilizing quantum interferences to decouple polarization entanglement from temporal correlation, we improve multi-photon entanglement f\textcompwordmark idelity from $(58.7\pm 2.2)\%$ to $(75.5\pm 2.0)\%$. Our work paves the way to realize scalable and high-quality multi-photon states from quantum dots.
\end{abstract}
\keywords{Quantum Information, Single Quantum Dot, Entangled Photon Source, Microcavity, Quantum Interference}
\maketitle

\section{Introduction}

In the past two decades, optical quantum computation has made substantial progress, from the f\textcompwordmark irst demonstration of quantum teleportation \cite{bouwmeesterExperimentalQuantumTeleportation1997a} to recent prototypes of quantum computation using photons \cite{zhongQuantumComputationalAdvantage2020,zhongPhaseProgrammableGaussianBoson2021, madsenQuantumComputationalAdvantage2022}. However, scalable implementation of large-scale linear optical quantum computer remains a pursuing goal partly because of the lack of ideal photon sources. Non-linear optical processes, such as spontaneous parametric down conversion (SPDC) \cite{zhong12PhotonEntanglementScalable2018} and spontaneous four-wave mixing (SFWM) \cite{llewellynChiptochipQuantumTeleportation2020}, are the workhorses in the f\textcompwordmark ields of quantum computation, quantum communication and quantum metrology. But, an intrinsic drawback in these processes is probabilistic generation. 

A pivotal and competitive solution is using deterministic photon sources from solid state emitters \cite{aharonovichSolidstateSinglephotonEmitters2016}, especially the epitaxial III-V semiconductor quantum dots (QDs) \cite{senellartHighperformanceSemiconductorQuantumdot2017a}. Utilizing biexciton-exciton (XX-X) cascaded emission, QDs can deterministically generate polarization entangled photon pairs with near-unity f\textcompwordmark idelity \cite{huberStrainTunableGaAsQuantum2018a}. However, there are two main problems which remain to be solved before achieving scalable QD platform. One is the wavelength discrepancy among different QDs, which can be solved with external tuning techniques, such as electric field \cite{zhaiQuantumInterferenceIdentical2022}, strain field \cite{reindlPhononAssistedTwoPhotonInterference2017} or frequency conversion \cite{youQuantumInterferenceIndependent2021a}. The other is poor indistinguishability due to the inherent temporal correlation in cascaded emission process \cite{schollCruxUsingCascaded2020a}, which becomes more noticeable after other issues are solved. Although asymmetric Purcell enhancement is suggested to eliminate the temporal correlation, it has unbalanced ef\textcompwordmark f\textcompwordmark iciency for X and XX emissions which is a serious side-ef\textcompwordmark fect \cite{schollCruxUsingCascaded2020a}.

In this work, we propose a scheme to eliminate temporal correlation in QD entangled photon source by quantum interferences. We validate it experimentally through improved entanglement f\textcompwordmark idelity of the four-photon Greenberger–Horne–Zeilinger (GHZ) state. This scheme separates the polarization degree of freedom (DoF) from the temporal DoF, by passing X(XX) photons from two entangled photon pairs through polarizing beam splitters (PBSs) simultaneously to perform quantum interferences. With this, we can overcome the limitation of imperfect indistinguishability from the cascaded emission process.

\begin{figure*}
    \centering
    \includegraphics[width=1\textwidth]{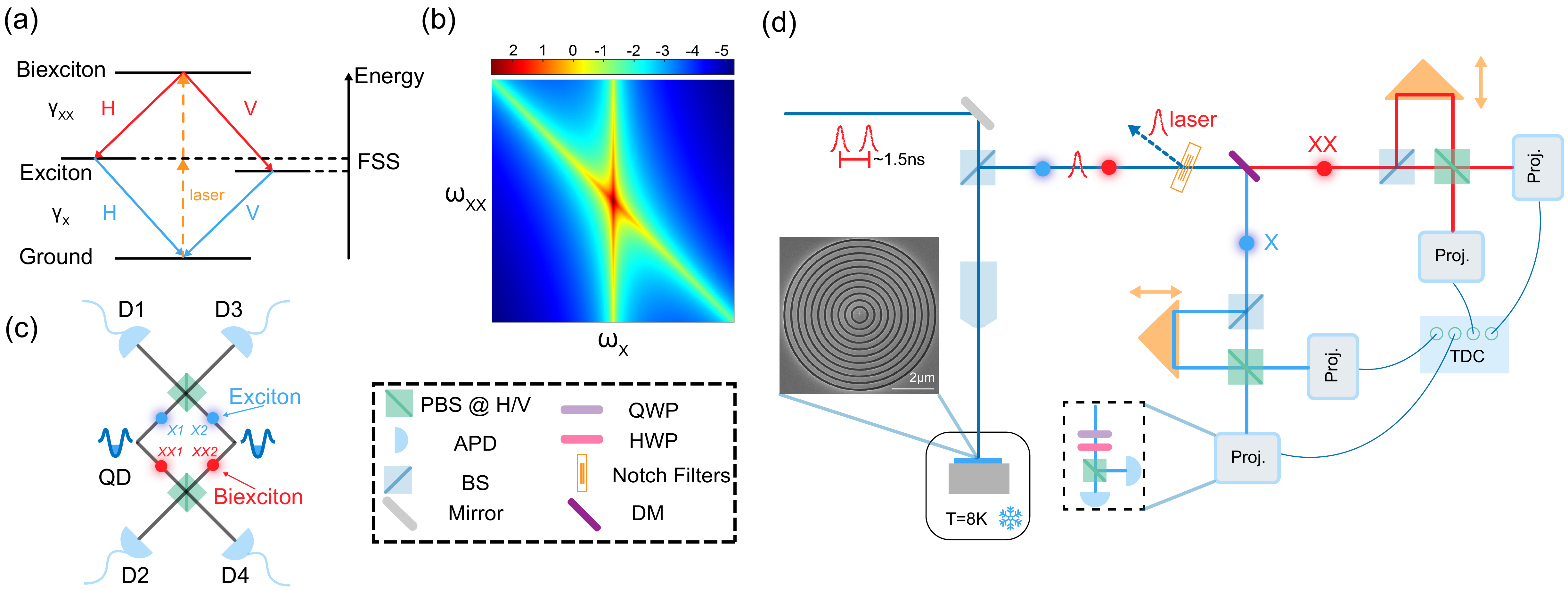}
    \caption{(color online) Schematics of principle and apparatus.
    \textbf{(a)} Diagram of energy levels for cascaded emission. 
    $\gamma_\text{XX}$ and $\gamma_\text{X}$ indicate the decay rates of XX and X, respectively.
    \textbf{(b)} Calculated joint probability of X and XX in the frequency domain, which is the Fourier transform of the temporal correlation. Here, $\gamma_\text{XX}=3.22\gamma_\text{X}$, deduced from the lifetime measurement of the investigated single QD.
    \textbf{(c)} Schematic of the experiment. Dif\textcompwordmark ferent from the usual single PBS to project independent entangled photon sources, we interfere X and XX photons together to eliminate temporal correlation.
    \textbf{(d)} Experimental setup. See the maintext for details. \iffalse A GaAs/AlGaAs QD embedded in a bull's-eye microcavity consecutively emits two entangled photon pairs. Then notch f\textcompwordmark ilter suppresses the excitation laser and DM routes X and XX to the ref\textcompwordmark lection and transmission paths. Beam splitters are used to de-multiplex photons to unbalanced Mach-Zehnder interferometers and impinge on PBSs for quantum interference. After that, photons are projected to arbitrary basis using a quarter waveplate (QWP), half waveplate (HWP) and PBS, and then, measured in APDs. Correlation events are analyzed by the TDC module.\fi }
    \label{fig:idea}
\end{figure*}

F\textcompwordmark irst, we brief\textcompwordmark ly summarize the temporal correlation in QD cascaded emission. As shown in \figref{fig:idea}a, the biexciton state can be deterministically prepared by two-photon excitation. For the exciton state, f\textcompwordmark ine structure splitting (FSS) is introduced because of the conf\textcompwordmark inement potential anisotropy originating from random QD growth \cite{bayerFineStructureNeutral2002}. Entangled photon pair can be generated radiatively from XX-X cascaded emission when the FSS is small enough to avoid which-way information leakage, with decay rates of $\gamma_\text{XX}$ and $\gamma_\text{X}$ for XX and X respectively. In this scenario, two-photon wavefunction in the time domain is illustrated as \cite{simonCreatingSingleTimeBinEntangled2005b,schollCruxUsingCascaded2020a}

\begin{align}
    \psi(t_1,t_2)=2\sqrt{\gamma_\text{XX}\gamma_{\text{X}}}e^{-\gamma_\text{XX}t_1}\text{H}(t_1)e^{-\gamma_\text{X}(t_2-t_1)}\text{H}(t_2-t_1)
\end{align}
where, $t_1$ and $t_2$ are the emission times of XX and X photons, respectively. $\text{H}(t)$ is the Heaviside function. The temporal correlation lies in the fact that the emission time of X photon is def\textcompwordmark initely later than the emission time of XX photon. This correlation is more intuitive in the frequency domain, as shown in \figref{fig:idea}b. The joint spectrum is 

\begin{align}
    &\psi(\omega_\text{XX},\omega_\text{X})=\frac{\gamma_\text{XX}\gamma_\text{X}}{\pi^2(\omega_\text{X}^2+\gamma_\text{X}^2)\left[(\omega_\text{X}+\omega_\text{XX})^2+\gamma_\text{XX}^2\right]}
\end{align}
It is apparent that the joint spectrum is non-separable. When a subsystem (X or XX) is considered, the quantum state of the corresponding photon is not pure. The degraded indistinguishability, which equals to $\gamma_\text{XX}/(\gamma_\text{XX}+\gamma_\text{X})$,  restricts the performance of multi-photon experiments in previous entanglement swapping demonstrations \cite{bassobassetEntanglementSwappingPhotons2019,zopfEntanglementSwappingSemiconductorGenerated2019a}.

\section{Experiments Configuration}

To solve this problem, an intuitive solution is to increase the value of $\gamma_\text{XX}/\gamma_\text{X}$, for example, by setting XX in a stronger Purcell regime than X in an optical cavity. However, this will result in different extraction ef\textcompwordmark f\textcompwordmark iciencies for XX and X, which hinders further applications. To avoid this dilemma, in this work, we propose a scheme to isolate the polarization DoF from temporal DoF. Thus, temporal correlation will not af\textcompwordmark fect polarization entanglement. \figref{fig:idea}c shows the schematic of our experiment. Compared with the generally used single PBS fusion operation \cite{bodiyaScalableGenerationGraphState2006a} to 
concatenate two pairs of independent entangled photons, we exploit quantum interferences for X photons and XX photons together. In this situation, because detectors D1\&D2 or D3\&D4 always respond to entangled photons from the same source, polarization entanglement is isolated from temporal correlation (see Supplementary Material for derivation). Thus, we can eliminate temporal correlation in QD entangled photon source and extend it to multi-photon experiments. 

\subsection{Apparatus}

Our experimental apparatus is shown in \figref{fig:idea}d. A single quantum dot is resonantly excited by a pulsed laser with a duration of $\sim$10 ps from a home-built 4f-pulse-shaper. Two entangled photon pairs emit consecutively, with a time delay of $\sim$1.5 ns, def\textcompwordmark ined by the delay line in the excitation path. A set of notch f\textcompwordmark ilters is used to suppress the excitation laser and preserve the quantum properties of emissions. Due to the energy discrepancy between X and XX photons, a suitable dichroic mirror (DM) is used to separate them in path DoF. In our case, X emission is in the ref\textcompwordmark lection path, and XX emission is in the transmission path. After that, two sequential X and XX photons are passively de-multiplexed to two arms of unbalanced Mach-Zehnder interferometer to compensate time dif\textcompwordmark ference, and then, interfere at PBSs. F\textcompwordmark inally, a set of quarter waveplate (QWP) and half waveplate (HWP) are used to choose the projection basis,  and photons are measured by PBS and avalanche photodiodes (APDs). We use a time-to-digital converter (TDC) to record the coincidences and analyze the correlation of multi-photon events.

\begin{figure*}
    \centering
    \includegraphics[width=\textwidth]{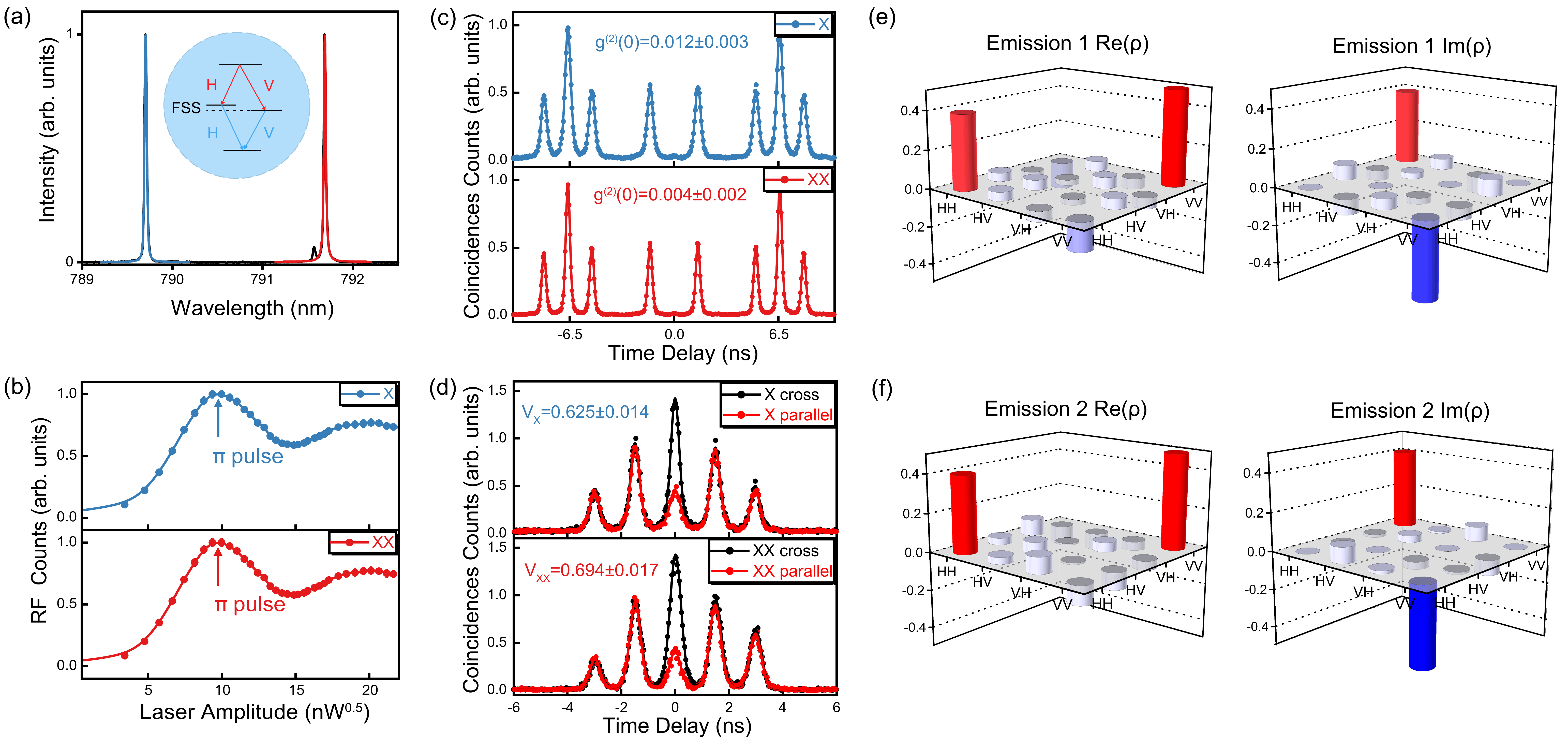}
    \caption{(color online) Systematic characterizations of selected QD.
    \textbf{(a)} Resonant f\textcompwordmark luorescence spectra of X and XX. The inset shows the energy levels diagram.
    \textbf{(b)} Rabi oscillations indicate that the excitation laser is coherently driving the QD. $\pi$ pulse power is $\sim$90 nW.
    \textbf{(c)} Second-order correlation of X(XX) using the HBT setup. We can deduce that $g^{(2)}_\text{X}(0)=0.012\pm0.003$ and $g^{(2)}_\text{XX}(0)=0.004\pm0.002$.
    \textbf{(d)} Two photon interference of X(XX) using PBS-type unbalanced Mach-Zehnder interferometer. The raw visibility is $0.625\pm0.014$ for X and $0.694\pm0.017$ for XX. Compared with the temporal correlation def\textcompwordmark ined upper bound indistinguishability of 0.76, QD is also af\textcompwordmark fected by spectral wandering.
    \textbf{(e, f)} Quantum tomography for consecutively emitted entangled photon pairs. The f\textcompwordmark idelity of state to a maximally entangled state is $\mathcal{F}=90.8\%\pm0.5\%$ and $\mathcal{F}=90.7\%\pm0.5\%$.}
    \label{fig:QD}
\end{figure*}

\subsection{QD Entangled Photon Source}

We f\textcompwordmark irstly characterize the dif\textcompwordmark ferent properties of single QD to determine the quality of entangled photon pairs. In the experiment, we employ GaAs/AlGaAs QD embedded in a bull's-eye optical microcavity \cite{wangOnDemandSemiconductorSource2019a,liuSolidstateSourceStrongly2019a}, grown by local droplet etching using P700 Molecular Beam Epitaxy (MBE) system \cite{huoUltrasmallExcitonicFine2013a,huoLightholeExcitonQuantum2014a,huoVolumeDependenceExcitonic2014}. Due to the symmetric shape of our QDs, whose FSS are mostly smaller than $4 \mu$eV, we can easily f\textcompwordmark ind a QD to emit highly entangled photon pairs. In addition, the Purcell effect of the microcavity can decrease the lifetimes of XX and X, and further enhance the possibility of achieving higher entanglement f\textcompwordmark idelity. \figref{fig:QD}a shows the resonant f\textcompwordmark luorescence spectrum of the investigated QD. Two-photon resonant excitation is used to coherently drive the XX-X system. The XX is deterministically excited when the energy of the laser is located at $(E_\text{X}+E_\text{XX})/2$, where $E_\text{X} (E_\text{XX})$ is the X(XX) energy. We vary the excitation laser power and record the resonant f\textcompwordmark luorescence counts in APD. As shown in \figref{fig:QD}b, a clear Rabi oscillation can be observed, and the $\pi$ pulse power is $\sim$90 nW. The maximum count rate is $\sim$2.4 M/s at APD (see Supplementary Material for ef\textcompwordmark f\textcompwordmark iciency analysis). At $\pi$ power, we examine the second-order correlation using a Hanbury-Brown-Twiss (HBT) interferometer, as shown in \figref{fig:QD}c. A nearly vanishing coincidence can be observed at zero time delay. We extract the $g^{(2)}_\text{X}(0)=0.012\pm0.003$ and $g^{(2)}_\text{XX}(0)=0.004\pm0.002$ without any background subtraction. 

Indistinguishability of single photons, which describes the overlap of wavefunctions, is one of the most important f\textcompwordmark igures of merit in multi-photon experiments. Here, we use an unbalanced Mach-Zehnder interferometer to measure visibility, which is, in a negligible multi-photon emission situation, the same as the indistinguishability \cite{fischerParticleEmissionOpen2018}. Dif\textcompwordmark ferent from the regular beam-splitter (BS) type interferometer, here, we employ a PBS type Hong-Ou-Mandel (HOM) interferometer. Single photons from X (or XX) are initialized by a polarizer in $|\text{D}\rangle=1/\sqrt{2}(|\text{H}\rangle+|\text{V}\rangle)$. Then, we measure the PBS outputs correlation at $|\text{DD}\rangle$ (parallel) and $|\text{DA}\rangle$ (cross) bases, as shown in \figref{fig:QD}d. The raw visibility is $0.625\pm0.014$ for X and $0.694\pm0.017$ for XX. We also measure the lifetimes of photons (see Supplementary Material for lifetime measurement), that is, $\sim125.5$ ps for X and $\sim38.8$ ps for XX. According to $\gamma_\text{XX}/(\gamma_\text{XX}+\gamma_\text{X})$ \cite{schollCruxUsingCascaded2020a}, for this QD, the upper bound of indistinguishability is $0.76$, which is larger than the measured values. This is because spectral wandering processes degrade the coherence of photons further \cite{kambsLimitationsIndistinguishabilityPhotons2018}.




The last important property of an entangled photon source is the f\textcompwordmark idelity of entanglement. The quantum tomographies \cite{jamesMeasurementQubits2001} for early (0 ns) and late (1.5 ns) photon pairs are illustrated in \figref{fig:QD}e and \figref{fig:QD}f, respectively. It is obvious that there is a phase shift between $|\text{HH}\rangle$ and $|\text{VV}\rangle$, which is from the birefringence of optical components. A direct calculation of f\textcompwordmark idelity to four Bell states is not applicable, so, here we calculate the maximum f\textcompwordmark idelity to the maximally entangled state, as described in Ref.  \cite{verstraeteFidelityMixedStates2002}. The f\textcompwordmark idelities for early and late photon pairs are $\mathcal{F}=90.8\%\pm0.5\%$ and $\mathcal{F}=90.7\%\pm0.5\%$, respectively.

\begin{figure}
    \centering
    \includegraphics[width=1\columnwidth]{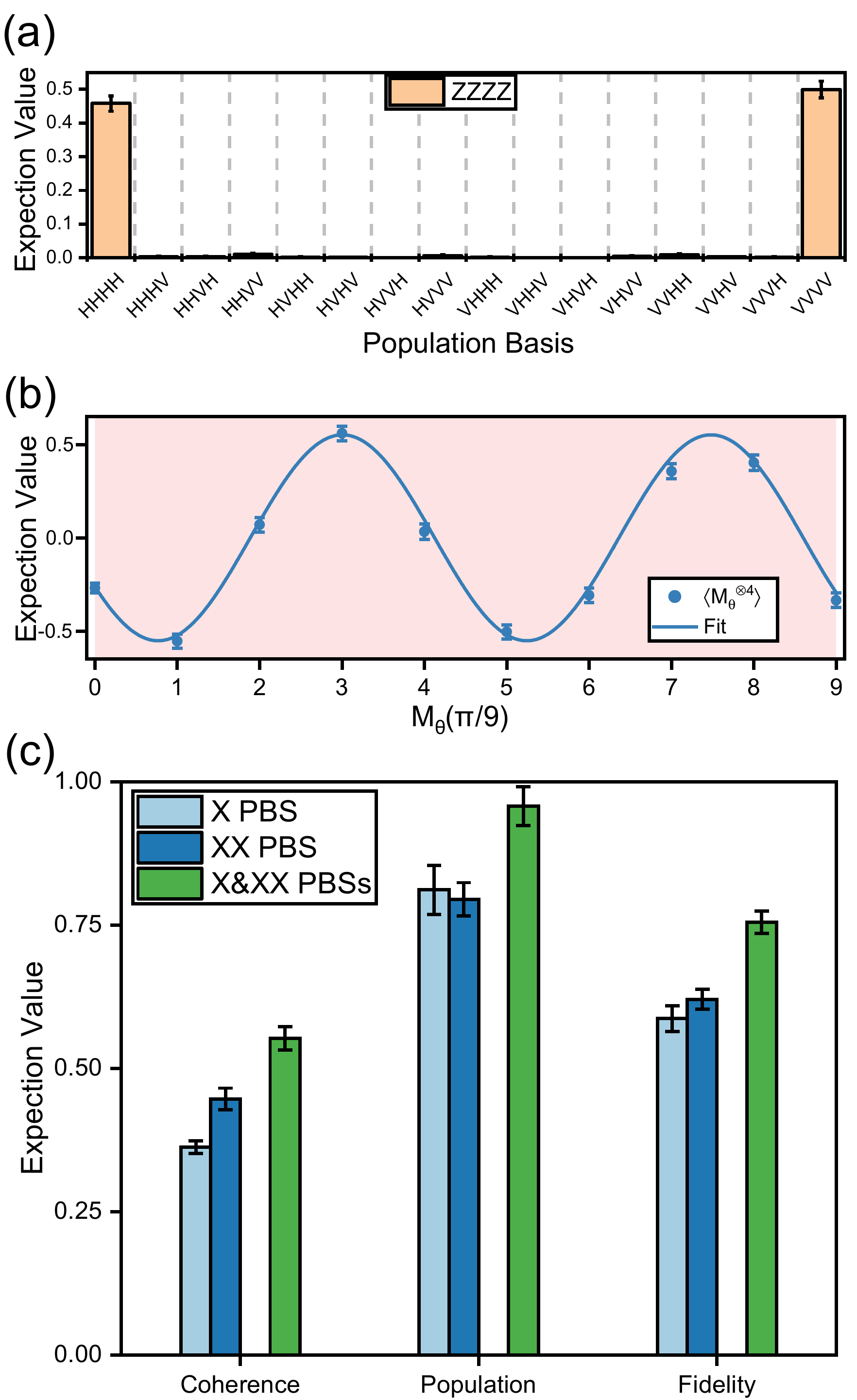}
    \caption{(color online) Experimental results. When X and XX interference occur simultaneously, 
    \textbf{(a)} Population of the four-photon GHZ state;
    \textbf{(b)} Coherence of the four-photon GHZ state. 
    \textbf{(c)} Comparison of coherence, population and f\textcompwordmark idelity for two PBS fusion operations and a single PBS fusion operation. When only one PBS is used for concatenating entangled photon pairs, the coherence is $0.362\pm0.011$ and $0.446\pm0.019$ for X and XX interference, respectively. When two PBSs are applied, the coherence is improved to $0.552\pm0.020$, which exceeds the single PBS situation by more than 3.83 standard deviations and 8.28 standard deviations for XX and X interference, respectively.}
    \label{fig:GHZ}
\end{figure}

\section{Results and Discussions}

After full characterizations of investigated QD, we turn to validate our scheme in multi-photon experiment, here, four-photon GHZ state generation. The most direct evidence is that the GHZ state f\textcompwordmark idelity can be improved when quantum interferences are used to decouple polarization entanglement from temporal correlation. Generally, considering the relative phase $\phi$ between orthogonal polarizations, the GHZ state for $N$-qubit is $|\Psi\rangle=1/ \sqrt{2}(|\text{H}\rangle^{\otimes N}+e^{i\phi}|\text{V}\rangle^{\otimes N})$. In density matrix formalism \cite{guhneToolboxEntanglementDetection2007},

\begin{subequations}
\begin{align}
    \rho_\text{\tiny GHZ}=&\frac{1}{2}(|\text{H}\rangle\langle\text{H}|^{\otimes N}+|\text{V}\rangle\langle\text{V}|^{\otimes N})\label{eqn:polulation}\\
    &+\frac{1}{2N}\sum_{i=0}^{N-1}(-1)^i\left[\mathcal{M}\left(\frac{i\pi}{N}-\frac{\phi}{N}\right)\right]^{\otimes N}\label{eqn:coherence}
\end{align}
\end{subequations}
where, $\mathcal{M}(\theta)=\cos\theta\sigma_x+\sin\theta\sigma_y$. The f\textcompwordmark idelity of the GHZ state can be measured by population $\langle P^{N}\rangle$ and coherence $\langle C^{N}\rangle$, as $\mathcal{F}=1/2(\langle P^{N}\rangle+\langle C^{N}\rangle)$. Population corresponds to diagonal elements in the density matrix, and coherence denotes the of\textcompwordmark f-diagonal elements in the density matrix. 

As shown in \figref{fig:GHZ}a, the population is measured by four-photon coincidence when the projection basis is H/V. $|\text{H}^{\otimes 4}\rangle$ and $|\text{V}^{\otimes 4}\rangle$ dominate the probability distribution, and we can conclude that $\langle P^N\rangle=0.956\pm0.034$. Compared with population, which ref\textcompwordmark lects the classical combination of dif\textcompwordmark ferent polarizations, coherence represents the quantum superposition between $|\text{H}^{\otimes 4}\rangle$ and $|\text{V}^{\otimes 4}\rangle$. In \figref{fig:GHZ}b, we measure the expectation values of $\mathcal{M}(\theta)$, where $\theta=i\pi/9, i=0,1,\cdots,9$. Sine function is used to f\textcompwordmark it and extract the coherence value as $\langle C^N\rangle=0.552\pm0.020$. As a result, the f\textcompwordmark idelity of the four-photon GHZ state with eliminated temporal correlation is $\mathcal{F}=0.755\pm0.020$. We note that in coherence measurement, there is an observable phase between orthogonal polarizations, which is from accumulated birefringence in optical paths and can be compensated with a Soleil-Babinet compensator. During the measurement, the coincidence window is set to 600 ps, which is approximately 5 times broader than the X lifetime. Therefore, clearly, there is no time post-selection. We also measure the coherence and population of a single PBS fusion operation for X or XX photons, and summarize the values in \figref{fig:GHZ}c. Obviously, when simultaneous quantum interferences are applied, coherence and population both increase, as a result, the f\textcompwordmark idelity improves. Actually, population improvement is a trivial effect when two PBSs are used. As the PBS only transmits $H$ and ref\textcompwordmark lects $V$ polarization, only two photons with the same polarization contribute to four-photon coincidence. Using two PBSs, residual $|\text{HV}\rangle$ and $|\text{VH}\rangle$ components in XX-X entanglement are f\textcompwordmark iltered.

In contrast, the coherence improvement is the direct evidence of eliminated temporal correlation. When only one PBS is used to concatenate two entangled photon pairs, the coherence values are $0.362\pm 0.011$ and $0.446\pm 0.019$ for X interference and XX interference, respectively. When two PBSs are simultaneously used for quantum interferences, coherence value is improved to $0.552\pm 0.020$, which exceeds the single PBS situation by more than 3.83 standard deviations and 8.28 standard deviations for XX and X interference, respectively. This improvement proves that, in the view of quantum state engineering, when separating the polarization DoF from temporal DoF, the left polarization entanglement is a pure state.

We point out that our proposal promises to improve the multi-photon GHZ state fidelity to be unity for cascaded-emission emitters. However, in experiments, multi-photon state entanglement f\textcompwordmark idelity is also limited by spectral wandering of photons and imperfect initial XX-X entanglement (see Supplementary Material for the theoretical model). This strongly indicates that transform-limited and near-perfect QD entangled photon sources are eager in future experiments. Embedding QDs in n-i-p diode can suppress charge noise and blinking phenomena \cite{zhaiLownoiseGaAsQuantum2020, kuhlmannChargeNoiseSpin2013a}. Precise cavity-emitter coupling \cite{tommBrightFastSource2021,najerGatedQuantumDot2019a} can amend the adverse ef\textcompwordmark fects of dephasing processes in solid-state environment and induce the Purcell ef\textcompwordmark fect to enhance extraction ef\textcompwordmark f\textcompwordmark iciency. Integrating QDs with piezo-electric actuator can alleviate FSS-induced degradation to f\textcompwordmark idelity and concurrence of entangled photon pairs \cite{huberStrainTunableGaAsQuantum2018a}.

\begin{figure}
    \centering
    \includegraphics[width=\columnwidth]{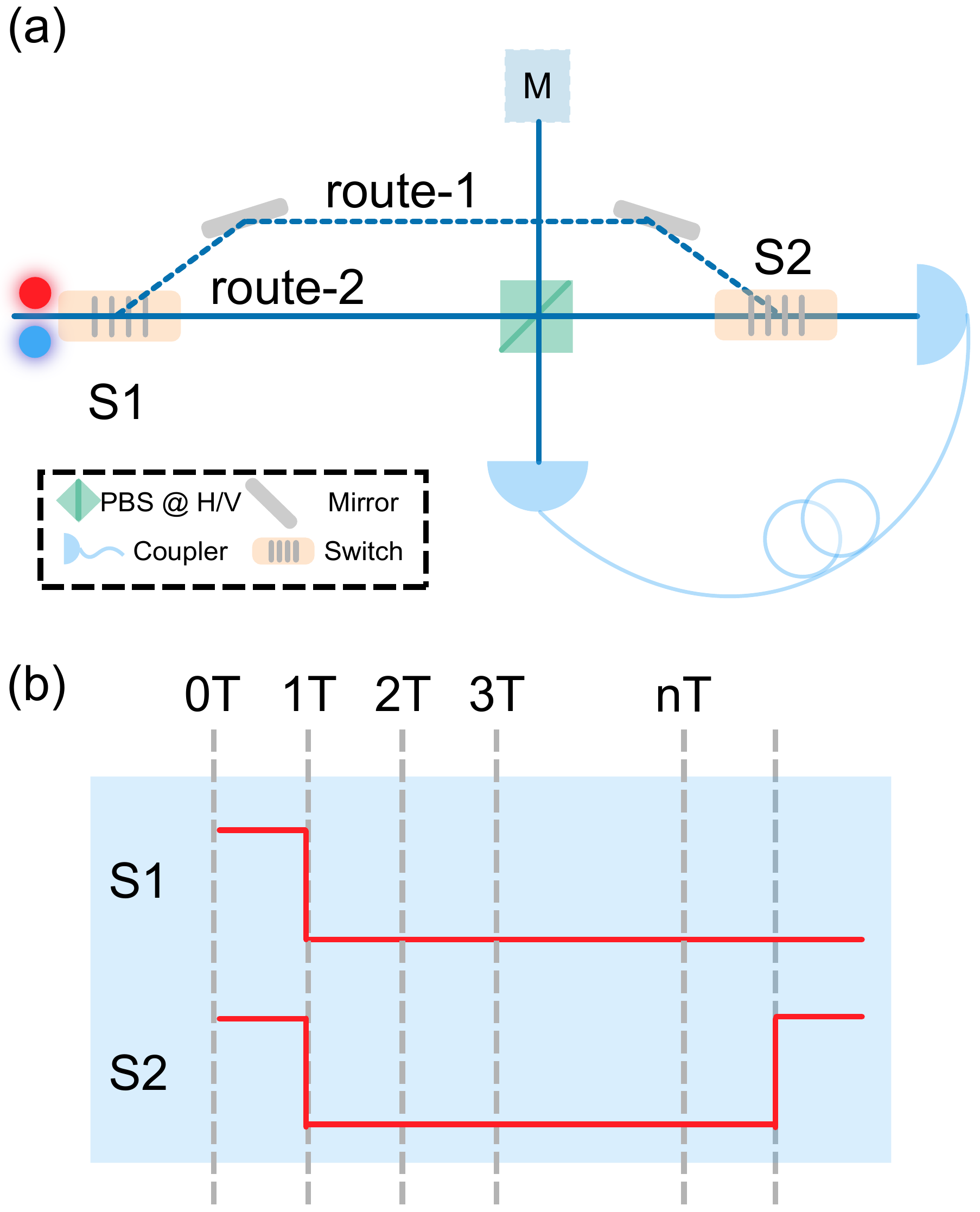}
    \caption{(color online) Schematic of the protocol extended to multi-photon GHZ generation. 
    \textbf{(a)} Two optical switches guide a pair of interfered photons to the input port of the PBS, then, these photons interfere with the next emission from the QD. F\textcompwordmark inally, multi-photon GHZ state can be addressed in the time domain. M denotes measurement setup, including single photon detector, TDC, suitable data processing, etc.
    \textbf{(b)} Signals of optical switches. In the f\textcompwordmark irst period, two optical switches can be opened to guide the f\textcompwordmark irst pair of entangled photon to pass through the f\textcompwordmark iber loop. Then, the multi-photon GHZ state can be generated by sequential quantum interference in PBS.}
    \label{fig:multiphotons}
\end{figure}

Lastly, we propose a generalized apparatus for multi-photon generation using a deterministic QD entangled photon source, as shown in \figref{fig:multiphotons}a. Optical switches, controlled by signals in \figref{fig:multiphotons}b, guide the f\textcompwordmark irst photon pair along route-1 and the following photon pairs along route-2 into a fiber loop. The consecutive photon pairs interfere at the PBS, as shown in the figure. Then, one half of the interfered photons leave the apparatus, and the other half of the photons are stored in the f\textcompwordmark iber loop. F\textcompwordmark inally, the multi-photon GHZ state is addressable in time domain.

\section{Conclusion}

In summary, we have demonstrated that quantum interferences can be exploited to eliminate temporal correlation in QD entangled photon source. The temporal correlation is inherited from cascaded emission and excludes the X(XX) photon indistinguishability from being unity, hindering scalable quantum connection between QDs. With simultaneous quantum interferences for X and XX photons, we disentangle the polarization DoF from temporal correlation and prove the coherence and f\textcompwordmark idelity improvement for four-photon GHZ state generation. Our work makes an essential step towards interfacing mutual QDs. From application point of view, a large multi-partite GHZ state is ready to be prepared using a single QD or separated QDs. More prospectively, a constant-sized entangled photon state can be produced and, with the aid of a boosted type-II fusion operation \cite{griceArbitrarilyCompleteBellstate2011}, a prototype of fusion based quantum computation (FBQC) \cite{bartolucciFusionbasedQuantumComputation2021} can be achieved. With respect to quantum communications, direct applications include demonstration of quantum swapping and entanglement distribution between independent QDs. Scalable and coherent quantum states are also useful in building the interface between QDs and quantum memory \cite{sinclairSpectralMultiplexingScalable2014,guhaRatelossAnalysisEfficient2015,woltersSimpleAtomicQuantum2017a}.

\section*{Conflict of interest}
The authors declare that they have no conflict of interest.

\begin{acknowledgments}
We gratefully acknowledge f\textcompwordmark inancial support from National Natural Science Foundation of China (Grant No. 11774326), National Key R\&D Program of China (Grant No. 2017YFA0304301), 2030 Major Project (Grant No. 2021ZD0300204), Shanghai Municipal Science and Technology Major Project (Grant No.2019SHZDZX01) and Anhui Initiative in Quantum Information Technologies.
\end{acknowledgments}

\section*{Author Contributions}
Yong-Heng Huo, Chao-Yang Lu and Jian-Wei Pan conceived the research. Run-Ze Liu, Han-Sen Zhong, Hui Wang, Chao-Yang Lu, Yong-Heng Huo and Jian-Wei Pan designed the experiments. Run-Ze Liu and Yong-Heng Huo grew the quantum dot sample. Run-Ze Liu, Zhen-Xuan Ge and Yong-Heng Huo designed the parameters of microcavity. Tung-Hsun Chung, Zhen-Xuan Ge and Yong-Heng Huo performed the nanostructure fabrication. Run-Ze Liu and Yun-Kun Qiao performed the optical experiments, including QD characterization and entangled state measurements. Run-Ze Liu, Yong-Heng Huo and Chao-Yang Lu analyzed the data and wrote the manuscript with input from all authors. Yong-Heng Huo and Jian-Wei Pan supervised the project.


%

\newpage
\clearpage

\end{document}